\documentclass[aps,prb,twocolumn,groupedaddress,showpacs]{revtex4}

\usepackage{graphicx}

\begin{document}

\title{Observation of Raman $G$-band splitting in  top-doped few-layer graphene}

\author{Matteo Bruna}
\author{Stefano Borini}
\email[]{s.borini@inrim.it} \affiliation{Electromagnetic Division,
INRIM, Strada delle Cacce 91, I-10135 Torino (Italy)}

\date{26 Feb 2010}

\begin{abstract}
An experimental study of Raman scattering in $N$-layer graphene as a
function of the top layer doping is reported. At high doping level,
achieved by a CHF$_3$ plasma treatment, we observe a splitting of
the $G$ band in the spectra of bilayer and 4-layer graphene ($N$
even), whereas the splitting is not visible in case of monolayer and
trilayer graphene ($N$ odd). The different behaviors are related to
distinct electron-phonon interactions, which are affected by
symmetry breaking and Fermi level position in different ways in the
various $N$-layer graphenes. In trilayer graphene, a weakening of
the electron-phonon coupling as a function of the Fermi energy
induces a hardening of all zone-center in-plane optical phonon
modes, like in monolayer graphene. On the other hand, in 4-layer
graphene two distinct trends are observed in the $G$ band as a
function of doping, suggesting the presence of two different groups
of electron-phonon interactions, like in bilayer graphene.

\end{abstract}

\maketitle

\section{Introduction}
Since the demonstration of the isolation of a single atomic plane of
graphite (graphene) on a standard $SiO_{2} / Si$ substrate
~\cite{Novoselov2005}, it has suddenly become possible to
experimentally verify many theoretical predictions about the
peculiar physical behavior of graphene-based systems. Indeed, the
linear $E-k$ dispersion curves, and the consequent relativistic-like
behavior of charge carriers in monolayer graphene, have been broadly
confirmed by several experimental observations
~\cite{CastroNeto2009,Geim2007}. Interestingly, stacking a number
$N$ of graphene layers on top of each other can lead to new physical
systems exhibiting completely different properties. For instance,
some peculiar gate-tunable electronic and optical properties have
been recently reported in the bilayer case
~\cite{Zhang2009,Mak2009b,Kuzmenko2009,Tang2010}.

In fact, the interlayer coupling induces a gradual departure from
the electronic bands of monolayer graphene ~\cite{Koshino2009},
until the bulk limit (graphite) is reached for $N$ large enough.
Therefore, there is a range of $N$ where the physical properties of
graphene stacks are sensitive even to a variation $\Delta N=1$. For
instance, a qualitative difference between the transport properties
of bilayer and trilayer graphene was evidenced in recent experiments
~\cite{Craciun2009}. Moreover, the optical absorption spectra are
predicted to systematically vary with the layers number within the
effective mass approximation for $1\leq N\leq 6$~\cite{Koshino2009},
reflecting the gradual modification of the band structure. Here we
show that a clear splitting of the Raman $G$ band is observed for
$N=2$ and $N=4$, when the multilayer graphene symmetry is broken by
heavy doping of the top layer, whereas the splitting is not observed
for $N=1$ and $N=3$. The presence or absence of splitting highlights
different electron-phonon interactions, which are influenced by
doping and symmetry breaking in distinct ways in the various
$N$-layer graphenes. The reported results confirm two experimental
reports in literature about the $G$ band splitting in bilayer
graphene (obtained by gate field effect)~\cite{Malard2008,Yan2009},
adding new important information such as a systematic study of the
splitting as a function of the layers number and of the doping.

Raman spectroscopy is a very powerful tool for studying graphene,
yielding information on the electronic structure and on the
electron-phonon coupling (EPC) in the material ~\cite{Ferrari2007,
Malard2009}. In fact, this technique allows to clearly distinguish a
monolayer and a bilayer from a few-layer graphene, through the
analysis of the $2D$ band ~\cite{Ferrari2006}, and to estimate the
charge carrier density and type in monolayer graphene from the
spectral positions and relative intensities of the $G$ and $2D$
bands ~\cite{Das2008}. The $G$ band (at $\sim 1580$ cm$^{-1}$) is
due to a first-order Raman scattering process involving zone-center
in-plane  optical phonons. In stacked graphene layers, the
vibrations in different atomic planes can combine with each other in
various ways, depending on the number of layers and on the stacking
order. Moreover, the EPC is affected by the number and symmetry of
the stacked layers, so that the $G$ band may be used to study the
effects of symmetry breaking and doping on the electronic and
phononic properties of multilayer graphenes. The knowledge of these
effects is of fundamental importance for the development of
graphene-based field-effect electronic devices.

\section{Experimental}
\begin{figure}
  \includegraphics{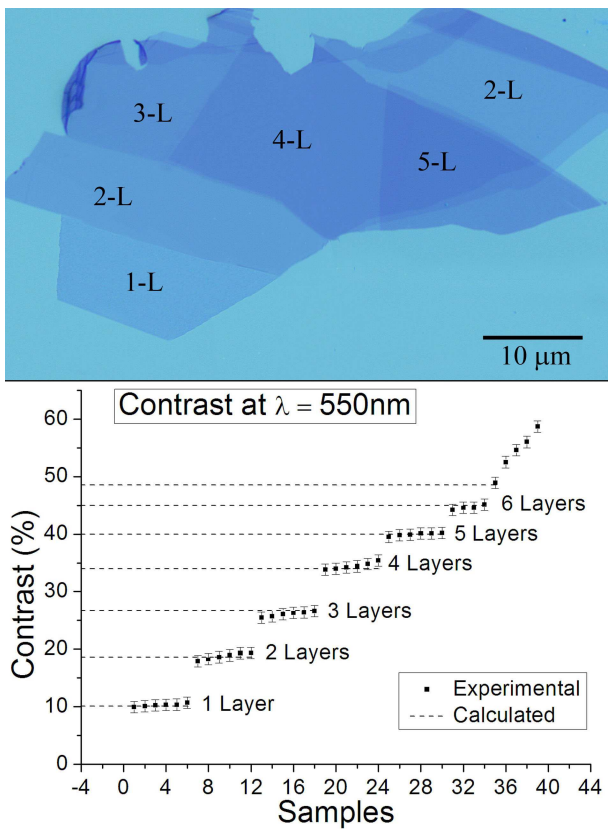}%
  \caption{(color online). Optical microscope image of one of the flakes analyzed in this work
(top). The number of layers estimated by contrast analysis is
indicated. In the bottom graph, the discrete behavior of the
contrast as a function of the number of layers is shown.
Experimental data, obtained at $\lambda$=550nm, are compared to
theoretical values.}\label{f1}
\end{figure}

Graphene layers studied in this work were deposited on a 285 nm
thick $SiO_{2}$ on $Si$ substrates, by adhesive tape exfoliation of
natural graphite. Then, the samples were analyzed by optical
microscopy, in order to estimate the number of graphene layers
composing the deposited thin flakes. It can be seen in Fig.~\ref{f1}
that the contrast (defined as $1-R_{G}/R_{S}$, where $R_{G}$ and
$R_{S}$ are the reflected light intensities from the $SiO_{2}/Si$
substrate with and without graphene, respectively) measured on many
semi-transparent flakes increases in a stepwise manner. The analysis
of the $2D$ Raman band confirmed that the lowest two steps
correspond to monolayer and bilayer graphenes, indicating that few
graphene layers can be counted by contrast analysis. Such a behavior
is related to the optical absorption of graphene, which was found to
be directly proportional to the number of layers for $N$ small
enough ~\cite{Nair2008}. Using appropriate filters in order to
select the wavelengths at which the contrast variation is high, we
were able to distinguish up to 6 layers. Moreover, the experimental
contrast values were checked by theoretical calculations within the
Fresnel coefficients approach~\cite{Bruna2009}.
After a preliminary annealing in vacuum (1x10$^{-5}$ mbar for two
days) in order to remove possible adsorbed impurities from the
graphene surface, the first run of Raman measurements was carried
out. Raman spectra were acquired by means of a Jobin-Yvon U1000
Raman spectrometer equipped with a microscope (100X objective) and
with an Ar-Kr laser, using the excitation wavelength $\lambda$ =
514.5 nm. The incident laser power focused on the sample was
adjusted to be less than 5 mW, to avoid any local heating effect.
Various $N$-layer graphene flakes were analyzed, displaying the
standard $G$ and $2D$ bands reported in literature.The $D$ band at
$\sim$ 1350 cm$^{-1}$, related to lattice defects, was never
observed in the experiments, confirming the good quality of our
graphene samples. Moreover, the analysis of the monolayer spectra
($G$ peak at $\sim$ 1582 cm$^{-1}$ with FWHM $\sim$ 13 cm$^{-1}$)
indicates that the unintentional doping level in the pristine
samples was relatively low (about 1X10$^{12}$
cm$^{-2}$)~\cite{Pisana2007}.

Then, the samples were subjected to a radio frequency (RF) plasma
treatment in CHF$_{3}$ gas, and immediately (within a few minutes)
re-characterized by Raman spectroscopy.

\section{Results and Discussion}

\subsection{CHF$_{3}$ plasma}

Previous studies reported in literature ~\cite{Tachibana1999,
Hancock2001} have shown that the following radical species can be
found in a CHF$_{3}$ plasma: $F$ atoms and $CF_{x}$ ($x=1,2,3$)
radicals. In dry etching processes, $F$ atoms normally act as the
reactive species (responsible for the etching), whereas $CF_{x}$
radicals are passivation precursors giving rise to polymer
deposition. It has been observed that, in the case of CHF$_{3}$
plasma processes, the etch rate decreases with increasing flow (in
contrast with what happens in the case of CF$_{4}$ plasma), due to
the low concentration ratio $[F]/[CF_{x}]$ ~\cite{Choi2008}. This
means that the action of $F$ atoms becomes more effective with
decreasing the flow rate, because the passivating action of $CF_{x}$
radicals is reduced. In our experiments, we have observed an
increasing modification of the graphene Raman spectra, with
decreasing the flow rate in the plasma treatment.

\begin{figure}
  \includegraphics{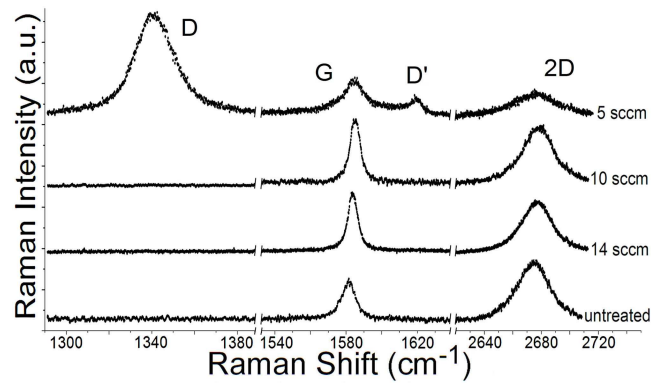}%
  \caption{Raman spectra of monolayer graphene after plasma treatments at various CHF$_{3}$ flow rates}\label{f2}
\end{figure}

In particular, we performed a preliminary study on monolayer
graphene processed at various CHF$_{3}$ flow rates. Fig.~\ref{f2}
shows that the plasma treatment induces a blue shift of both $G$ and
$2D$ peaks, which increases with decreasing the gas flow. Moreover,
below a flow rate threshold (about 6 SCCM), two new peaks arise at
about 1350 cm$^{-1}$ ($D$ peak) and 1620 cm$^{-1}$ ($D'$ peak),
which indicate the presence of defects in the $sp^{2}$ $C$ lattice.
Also the effect on the Raman spectra of $N$-layer graphenes, which
is going to be discussed in detail in the following Section, was
remarkably reduced with increasing the gas flow rate.

In analogy with the mechanism involved in the etching processes, the
interaction of $F$ atoms with the graphene surface is likely reduced
at high flow rate, because of the passivating action of the $CF_{x}$
radicals. Therefore, the effect of CHF$_{3}$ plasma on graphene
Raman spectra can be ascribed to the adsorption of $F$ atoms on the
surface. With increasing the $F$ coverage, the modification of the
graphene properties becomes more important, passing from a p-type
doping effect (blue-shifted $G$ and $2D$ peaks) ~\cite{Das2008} to a
structural modification ($D$ and $D'$ peaks) at a very low flow rate
(less than 6 SCCM). The doping effect can be ascribed to an electron
transfer from graphene to adsorbed $F$ atoms, i.e. a mechanism
analogous to the observed charge transfer between graphene and
adsorbed $K$ atoms ~\cite{Ohta2006}. At very low gas flow, when the
passivating action of the $CF_{x}$ radicals is minimized, chemical
modification (fluorination) of graphene may eventually occur, with a
transition from $sp^{2}$ to $sp^{3}$ $C$ hybridization similar to
that observed in graphane formation by plasma hydrogenation
~\cite{Elias2009}. This evolution may be analogous to the transition
from semi-ionic to covalent $C-F$ bonding observed in carbon
nanotubes treated in $CF_{4}$ plasma ~\cite{Plank2003}.

Here, we discuss the results of processes carried out at a gas flow
of 6 SCCM (pressure of 100 mTorr) for 5 minutes, at RF power = 15 W.
Such experimental conditions lead to a very high doping, without
structural modification of the graphenes. Moreover, the symmetry of
stacked graphenes is broken by the dipole moment generated by the
charge transfer from graphene to the adsorbed $F$ atoms. Therefore,
the situation under study is very similar to that found in
field-effect experiments, and the results here reported may be
useful for the study of gated graphene-based devices.

\subsection{$G$ band splitting and electron-phonon coupling in $N$-layer graphene}

\begin{figure*}
  \includegraphics{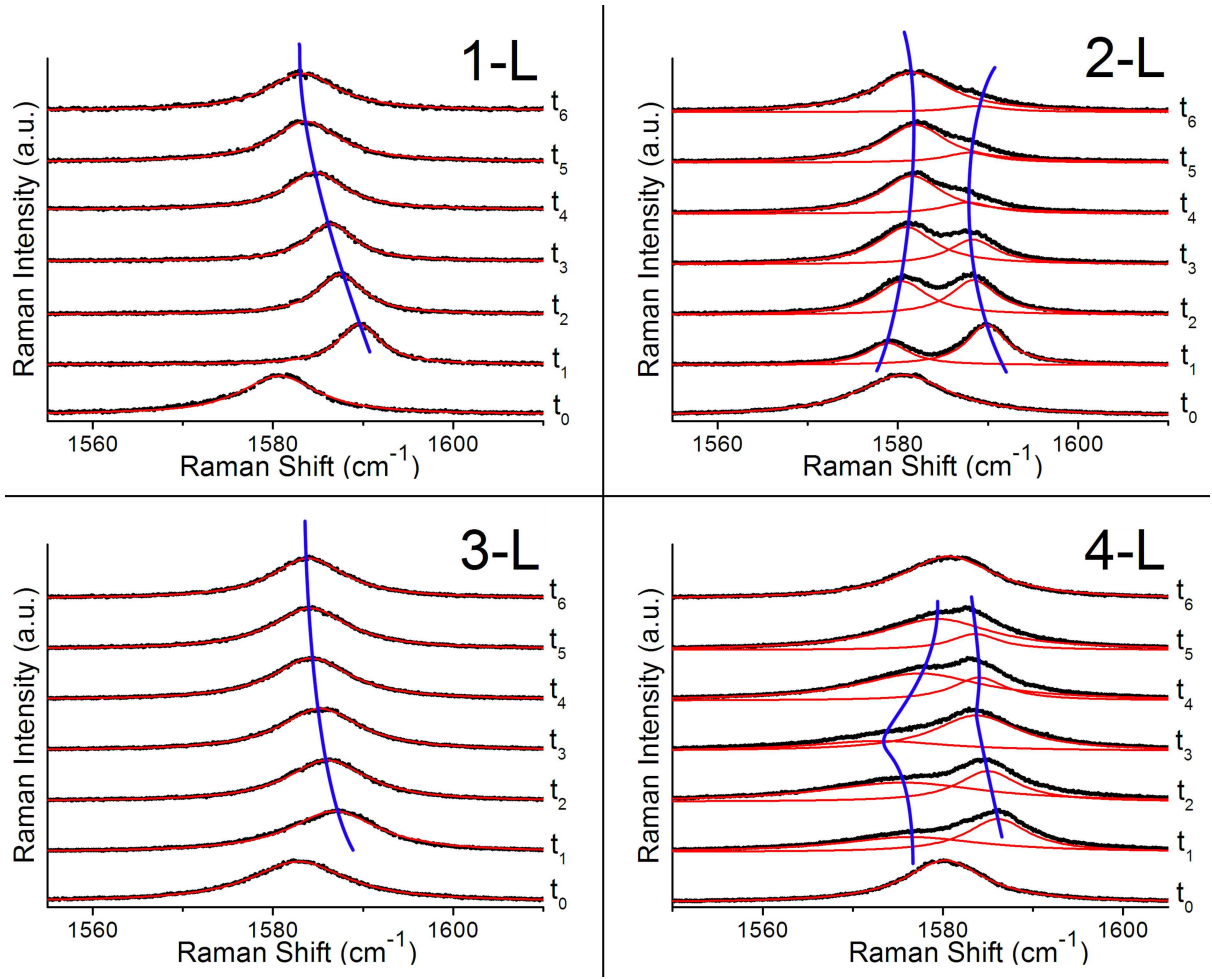}%
  \caption{(color online). Evolution of the Raman $G$ band of $N$-layer graphene after CHF$_{3}$ plasma treatment.
The spectra at $t_{0}$ were acquired before the treatment, while
those at $t_{i}$ ($1 \leq i \leq 6$) were taken at various time
intervals after the treatment ($t_{1}\sim 15 min$;$t_{2}\sim 6 h$;
$t_{3}\sim 24 h$; $t_{4}\sim 48 h$; $t_{5}\sim 72 h$; $t_{6}\sim 144
h$ ). The lines connecting peaks are guides for the eye.} \label{f3}
\end{figure*}

We focus now on the effect of top doping on the Raman $G$ band,
which displays very distinct features depending on the number $N$ of
stacked graphene layers. The change of the $G$ band induced by
plasma treatment in the various cases is visible in Fig.~\ref{f3},
where the spectra at $t_{0}$ and $t_{1}$ were acquired on the same
substrate before and immediately after the treatment, respectively.
The $G$ peak of monolayer graphene is largely blue-shifted (to
$\sim$ 1590 cm$^{-1}$) and narrowed (FWHM $\sim$ 6 cm$^{-1}$). Both
the observations are consistent with an increase of the doping
level, which induces a hardening of the mode, due to the
non-adiabatic removal of a Kohn anomaly for zone center optical
phonons ~\cite{Lazzeri2006}, and a reduction of the linewidth, due
to Pauli exclusion principle which inhibits phonon decay into
electron-hole pairs when the Fermi level surpasses half the phonon
energy ~\cite{Yan2007}. On the other hand, the bilayer and 4-layer
spectra display a very evident splitting of the $G$-mode, whereas
the behavior of trilayer spectrum is similar to that of monolayer.
These results have been confirmed on different flakes on the same
sample and on different samples.

Furthermore, we observed that the modification induced by the plasma
treatment was not stable under ambient conditions, as the Raman
spectra changed with the passing of time, slowly tending to their
pristine form (Fig.~\ref{f3}). Indeed, the initial condition can be
restored by a vacuum annealing, so that the plasma treatment can be
repeated for several times in a reproducible way. Such a reversible
behavior is consistent with the absence of the $D$ peak, which
indicates the lack of structural modification of the material (in
contrast with the case of the chemical modification obtained by
plasma treatment at very low gas flow). As previously discussed, the
monolayer spectra, showing the hardening of both the $G$-mode and
the $2D$ mode, tell us that a strong $p$-type doping is achieved
~\cite{Das2008} upon plasma processing. Therefore, we were able to
gradually vary the doping level on top of each $N$-layer graphene,
studying the effect for different values of $N$.

The effect of charged adsorbates has already been studied on
epitaxial bilayer graphene on SiC by ARPES measurements
~\cite{Ohta2006}. It was shown that the electronic bands of bilayer
graphene are strongly affected by a potassium atoms coverage on the
top surface, due to the $n$-type doping induced by the adsorbates.
Indeed, because of the short screening length along the $c$-axis of
stacked graphenes ~\cite{Ohta2007}, the majority of the doping
charge is localized in the top layer and a dipole moment is formed,
giving rise to a symmetry breaking and to the consequent band gap
opening ~\cite{McCann2006}. In our case, an analogous effect is
likely to occur: the doping charge density rapidly decreases going
from the top layer to the bottom, thus altering the symmetry of the
multilayers. The symmetry breaking can affect very much the physical
properties of the system, as it eliminates the symmetry constraints
which rule both phonon Raman activity and EPC. In these conditions,
all phonon modes included in the $G$ band can become Raman active
and strongly mixed with each other ~\cite{Ando2009}, and the EPC is
affected by the modification of the electronic bands.

In Fig.~\ref{f3}, the $G$ band dependence on the doping is visible
in all cases, but it looks different for different $N$ values. An
estimation of the doping level may be obtained from monolayer
graphene spectra, basing on the $G$ peak position and linewidth, and
on the intensity ratio of the $G$ and $2D$ peaks ~\cite{Das2008},
but we have verified that different monolayers can display slightly
different $G$ peak positions on the same sample, due to the
difficulty of controlling the native doping of graphene in ambient
atmosphere ~\cite{Casiraghi2007}. Furthermore, monolayer graphene is
likely to have a different reactivity with respect to that of
few-layer graphene, as experimentally observed in the case of
hydrogenation ~\cite{Elias2009,Ryu2008}. Also in our experiments, we
have seen that, after plasma treatment at very low flow rate, the
$D$ peak is hardly observed in the few-layers spectra, whereas the
monolayer spectrum displays a very evident $D$ peak. Therefore, it
seems not correct to extrapolate the doping values from the
monolayer analysis to interpret the few-layer spectra. However,
although a precise quantitative estimate of carrier concentration in
each case is not possible in our experiments, we can monitor the
different behavior of $N$-layer graphene Raman spectra with
decreasing the top doping, starting from high doping levels (more
than $1\times10^{13} cm^{-2}$) as suggested by the analysis of
monolayers spectra.

The effect in the bilayer can be interpreted on the basis of some
recent literature. Indeed, the splitting of the Raman $G$ band was
recently observed in gated bilayer graphene ~\cite{Malard2008}, and
ascribed to the inversion symmetry breaking induced by the gate
field effect and to two distinct EPC involved in the $G$ band. This
is due to the fact that the $G$ band of bilayer graphene includes
two doubly degenerate modes, $E_{2g}$ and $E_{u}$, which are
symmetric and antisymmetric with respect to the inversion symmetry,
respectively. Consequently, only the $E_{2g}$ mode is normally Raman
active, unless inversion symmetry breaking switches the $E_{u}$ mode
on, too. It has been shown by Ando and Koshino ~\cite{Ando2009}
that, in the presence of asymmetry in the potential of the two
stacked layers, symmetric and antisymmetric modes are strongly mixed
with each other and two peaks appear in the Raman spectrum.
Moreover, phonons can be considerably modified by resonant
electronic interband transitions, when the asymmetry opens up a gap
comparable to the phonon energy. Ab initio calculations have been
performed to compute the $G$ band of bilayer graphene in asymmetric
conditions as a function of the carrier concentration in the top and
bottom layer ($n_{top}$ and $n_{bot}$) ~\cite{Gava2009}, predicting
the behavior of the two modes which become Raman visible at certain
values of $n_{top}$ and $n_{bot}$.

\begin{figure}
  \includegraphics{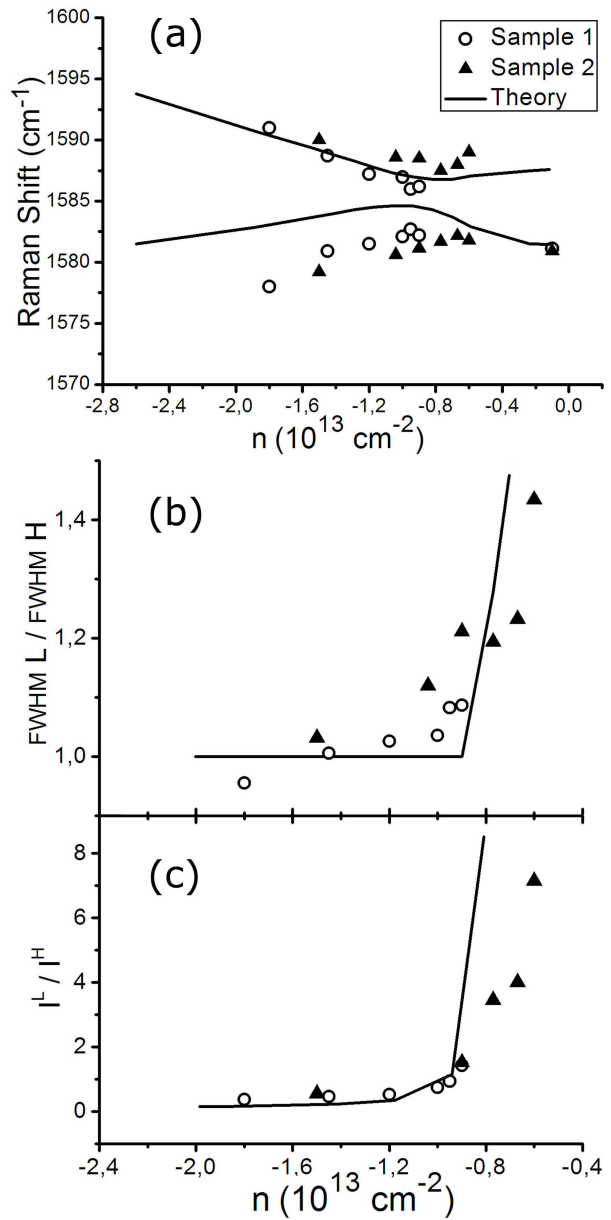}%
  \caption{Features of the two peaks fitting the bilayer $G$ band as a function of
  doping. Circles and triangles are experimental data from two
  different samples, and lines are the theoretical findings of
  Ref.~\cite{Gava2009}. In (a), the peaks positions
have been fitted by the theoretical curves, assuming that about the
$85\%$ of the total charge carriers is confined in the top layer. In
(b) and (c), the width and intensity ratios of the low (L) frequency
and high (H) frequency peak are reported.}\label{f4}
\end{figure}

In Fig.~\ref{f4}, we report the analysis of the two Lorentzian peaks
which can fit the bilayer $G$ band at various stages of the plasma
modification, for two different samples. The experimental evolution
of the peaks position has been fitted by the theoretical curves
obtained from ref.~\cite{Gava2009}, assuming that about the $85\%$
of the total charge carriers is confined in the top layer, according
to ref.~\cite{Ohta2007}, and using the total carrier concentration
$n$ as the free parameter in the fitting. Then, the other spectral
features (intensity and linewidth) have been compared to the
behavior predicted in ref.~\cite{Gava2009}, considering the carrier
concentration values obtained from the best fit of the peaks
position curve. It can be seen that a good qualitative agreement is
obtained for all the analyzed parameters. In particular, the
threshold at about $n = 1\times10^{13} cm^{-2}$, after which a steep
variation of both the intensity and linewidth ratios is predicted by
theory, is well reproduced by the experimental data, thus confirming
the consistence of the carrier concentration results obtained by
fitting the peaks position curves. Therefore, the behavior of the
bilayer $G$ band is well interpreted in the framework of an
asymmetric carrier distribution model.

In order to interpret the spectra for $N>2$, it is worth to
consider, first of all, the evolution of the $G$ band with $N$ as
predicted by group theory. Indeed, applying the group theory to
Bernal stacked graphene layers, the irreducible representations of
the infrared and Raman active modes at the $\Gamma$ point for
$N$-layer graphene can be obtained ~\cite{Jiang2008,Malard2009b}, as
listed in Table \ref{t1}. The modes related to the $G$ band are the
$E_{g}$, $E_{u}$, $E^{'}$ and $E^{''}$ in-plane modes, while the $A$
modes are related to out-of-plane phonons.

\begin{table}
  \centering
  \caption{Irreducible representations of the IR and Raman active modes at the $\Gamma$ point for $N$-layer graphene}\label{t1}
  \begin{tabular}{|c|c|c|}
  \hline
  $N$ & $\Gamma^{IR}$ & $\Gamma^{Raman}$ \\
  \hline
  Even & $(N-1)A_{2u}\oplus(N-1)E_{u}$ & $NA_{1g}\oplus NE_{g}$ \\
  Odd & $NA_{2}^{''} \oplus NE^{'}$ & $(N-1)A^{'}_{1} \oplus NE^{'} \oplus (N-1)E^{''}$ \\
  \hline
\end{tabular}
\end{table}

$E_{g}$ and $E_{u}$ modes are found in case of inversion symmetry of
the system ($N$ even), whereas $E^{'}$ and $E^{''}$ modes appear for
mirror symmetry ($N$ odd). Importantly, only the inversion symmetry
inhibits the Raman activity of antisymmetric modes. It can be
obtained by ab initio calculations ~\cite{Saha2008} that in trilayer
graphene two $E^{'}$ modes and one $E^{''}$ mode can be found at the
G-band frequencies, whereas in 4-layer graphene two $E_{g}$ modes
and two $E_{u}$ modes vibrate at the $G$ band frequencies. The
inversion symmetry breaking, produced by the top doping, switches on
the previously Raman silent antisymmetric modes $E_{u}$ in $N$
even-layer graphene, so that all the phonon modes become Raman
active and mixed in $N$-layer graphene for every $N$ value.
Moreover, like in the bilayer case previously discussed, the EPC can
be strongly affected by the lack of symmetry constraints and by the
change of the electronic band structures.

\begin{figure}
  \includegraphics{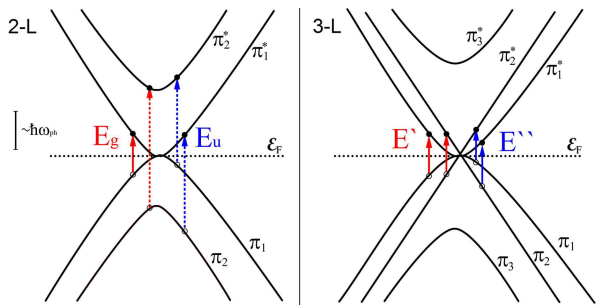}%
  \caption{(color online). Electronic transitions allowed by symmetry rules
in bilayer (left) and trilayer (right) graphene when
$\epsilon_{F}=0$.The electronic bands are taken from
Ref.~\cite{Koshino2009}. The transitions indicated by dotted lines
are not involved in the phonon energy renormalization, because their
energy is much higher than the $G$ band phonon energy
($\sim0.196eV$), which is reported as a scale bar on the
left.}\label{f5}
\end{figure}

The absence of splitting for trilayer graphene can be qualitatively
interpreted considering the allowed electronic interband transitions
which give rise to the phonon energy renormalization (Kohn anomaly)
when $E_{F} \sim 0$ (low doping). In Fig.~\ref{f5} it is shown that
both $E^{'}$ and $E^{''}$ phonons can couple with electronic
transitions when the Fermi level is at the Dirac point, whereas in
bilayer graphene only the symmetric $E_{g}$ mode can efficiently
create electron - hole pairs, due to energy conservation and
symmetry selection rules. Indeed, first principles calculations
~\cite{Yan2009b} have shown that in trilayer graphene the phonon
linewidths of symmetric and antisymmetric modes are all of the same
order of magnitude, whereas in the bilayer the antisymmetric
linewidth is two orders of magnitude smaller than that of the
symmetric mode. Therefore, almost the same EPC strength is expected
for the three phonon modes in trilayer graphene, giving rise to the
phonon energy renormalization at $E_{F}=0$ and to the consequent
hardening of all phonons with moving the Fermi level, like in
monolayer graphene. This is a qualitative different case with
respect to that of bilayer, where only the symmetric phonon energy
is renormalized at $E_{F}\sim0$. The presence or absence of the $G$
band splitting in the two cases reflects the presence or absence of
distinct EPC for phonons of distinct symmetry.

\begin{figure}
  \includegraphics{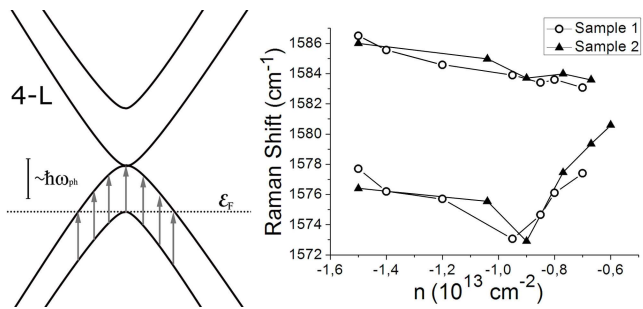}%
  \caption{(left) Electronic intraband transitions in 4-layer graphene
for $\epsilon_{F}=0.24 eV$. The electronic bands are taken from
Ref.~\cite{Koshino2009}, and the $G$ band phonon energy
($\sim0.196eV$) is reported as a scale bar. (right) Positions of the
two peaks fitting the splitted $G$ band of 4-layer graphene as a
function of doping. The observed dip may be related to a strong EPC
due to the intraband transitions shown on the left.}\label{f6}
\end{figure}

The observation of splitting in the 4-layer spectrum suggests the
presence of distinct EPC for the different phonon modes in analogy
with the bi-layer behavior. In Fig.~\ref{f6} we show the
experimental position of the two lorentzian peaks fitting the
4-layer $G$ band, as a function of doping as obtained from the
bilayer analysis. It is worth noting that a minimum can be clearly
identified in the curve of the low frequency peak, indicating that a
maximum is likely to occur in the EPC for some phonon modes at a
given doping value ($n_{m} \simeq 9X10^{12} cm^{-2}$).
Such a feature can be interpreted by taking into account the band
structure of the 4-layer, which can be approximated by two bilayer
type band structures ~\cite{Koshino2009}. In particular, intraband
transitions occurring between two sub-bands separated by $\Delta E
\sim 0.24 eV$, which corresponds to an experimentally observed
absorption by IR spectroscopy ~\cite{Mak2009}, may give rise to an
almost resonant coupling with phonons ($E_{ph}\sim 0.2 eV$), and to
a consequently strong renormalization of phonon energy. This EPC is
expected to have a maximum when the Fermi level reaches the high
energy sub-band (see Fig.~\ref{f6}), when the number of possible
intraband transitions is maximized. Therefore, the doping value
$n_m$, at which a minimum for the Raman peak position is observed,
is likely to correspond to the Fermi energy touching the high energy
sub-band at $E_{F}\sim 0.24 eV$. A theoretical analysis of the
4-layer band structure and density of states in the presence of
asymmetric doping may confirm this hypothesis.

\section{Conclusions}
In summary, we have experimentally investigated the Raman $G$ band
for $N$-layer graphene ($1 \leq N \leq 4$) in the presence of high
asymmetric doping, finding two different types of behavior. For $N$
odd, the $G$ band is always fitted by a single lorentzian peak,
which is blue-shifted with increasing the doping level. This is due
to a strong EPC for all phonon modes when $E_{F}\sim 0$, which
decreases with increasing the Fermi energy. On the other hand, for
$N$ even, an evident splitting of the $G$ band is observed, related
to the presence of distinct EPC for phonons of distinct symmetry. In
particular, in the 4-layer case a signature of the van Hove
singularity at $E \sim 0.24 eV$ is likely to be observed as a
minimum of the low energy peak position.

Insights into the electron - phonon interactions in $N$- layer
graphenes in the presence of top doping can be useful for the study
of field effect graphene-based devices. Moreover, the $CHF_{3}$
plasma treatment may be a powerful technique for the study of
graphene in the presence of a coverage of highly electronegative
atoms such as fluorine.

Finally, the variety of the results reported in the literature about
the Raman $G$ band in heavily doped bilayer graphene
~\cite{Malard2008,Yan2008,Yan2009,Das2009} suggests that the
repartition of the additional charge carriers is not well understood
in most experiments. Therefore, suspended graphene samples may be a
good testbed to further investigate the distribution of the doping,
especially in the bilayer case.~\cite{Berciaud2009,Ni2009}

\section{acknowledgments}
This work was carried out within the EURAMET Joint Research Project
"ULQHE". The research within this EURAMET JRP receives funding from
the EC FP7, ERA-NET Plus, under Grant Agreement No. 217257.

\end{document}